\documentclass[aps,pra,twocolumn,showpacs,floatfix,superscriptaddress,amsmath,amssymb,longbibliography,10pt]{revtex4-2}

\usepackage{graphicx}
\usepackage{dcolumn}
\usepackage{bm}
\usepackage{braket}
\usepackage{amssymb}
\usepackage{amsmath}
\usepackage{float}
\usepackage{multirow}
\usepackage{tgtermes}
\usepackage{ulem}
\usepackage{color}
\usepackage[english]{babel}
\usepackage{txfonts}
\usepackage[altg]{qtxmath}
\usepackage{microtype}
\usepackage{slashed}
\usepackage{appendix}

\makeatletter
\newcommand*{\rom}[1]{\expandafter\@slowromancap\romannumeral #1@}
\makeatother

\begin{document}

\author{E. Raicher}
\email{erez.raicher@mail.huji.ac.il }
\affiliation{Racah Institute of Physics, The Hebrew University, Jerusalem 9190401, Israel}
\author{Q.Z. Lv}
\email{qzlv@gscaep.ac.cn}
\affiliation{Graduate School, China Academy of Engineering Physics, Beijing 100193, China}

%\title{ Approximated solution to the Dirac equation of a relativistic electron in counterpropagating laser beams far from the resonance regime}

%\title{Quantum dynamics of a relativistic electron in counterpropagating laser beams far from the resonance regime}

\title{Nonresonant quantum dynamics of a relativistic electron in counterpropagating laser beams}

\begin{abstract}
  The quantum dynamics of an ultrarelativistic electron in strong counterpropagating laser beams is investigated.  In contrast to the stimulated Compton scattering regime, we consider the case when in the electron rest frame the frequency of the counterpropagating wave greatly exceeds that of the copropagating one. Taking advantage of the corresponding approximation recently developed by the authors for treating the classical dynamics in this setup, we solve the Klein-Gordon and Dirac equations in the quasiclassical approximation in the  Wentzel-Kramers-Brillouin (WKB) framework. The obtained wave function for the Dirac equation shows entanglement between the kinetic momentum and spin of the electron, while the spin-averaged momentum coincides with the classical counterpart as expected. The derived wave functions will enable the calculation of the probabilities of nonlinear QED processes in this counterpropagating fields configuration.
  %The starting point is an approximated solution to the classical equation of motion, recently obtained by the authors. Taking advantage of the classical trajectory, the Hamilton-Jacoby equation is solved, yielding the classical action. It is afterwards utilized in order to solve the Dirac and Klein-Gordon equations in the framework of the  Wentzel-Kramers-Brillouin (WKB) approximation.   Employing the wave function for the Dirac equation, we studied the spin effects on electron's dynamics and showed that the expectation value of the momentum operator depends on the spin direction, even the spin averaged momentum conincide with the classical counterpart. This new wave function can also enable the calculation of the probabilities of nonlinear QED processes in this counterpropagating field configuration.
\end{abstract}

\maketitle

\section{Introduction}

Owing to the rapidly developing laser technology, ultraintense coherent laser beams with intensity up to $\sim 10^{23}$~W/cm$^2$ can be generated \cite{Yoon_2021,Danson_2019,Radier_2022,Li_2022,Vulcan1}, corresponding to the normalized field amplitude $\xi \sim 300$, where
% (with the definition
$\xi = e\sqrt{-A^2}/m$, $m,e$ are the electron charge and mass respectively, and $A_{\mu}$ the laser vector potential. Relativistic units with $\hbar=c=1$ are used throughout. Since $\xi \gg 1$, the electron dynamics in the presence of the field is highly nonlinear and involve multiphoton emission and absorption. For ultrarelativistic electrons quantum effects in radiative processes are important when
%In case the particle moving in the field is energetic enough,
the field amplitude  $E'$ experienced in its rest frame approaches the Schwinger field $E_s=m^2/e$ \cite{Ritus_1985}. The ratio between these quantities is known as the quantum strong field parameter
$\chi \equiv E'/E_s = e\sqrt{-(F^{\mu \nu} P_{\nu})^2}/m^3$, where $F_{\mu \nu} $ is the electromagnetic tensor and $P_{\mu}=(\varepsilon,\mathbf{P})$ the particle 4-momentum.
% and relativistic units $\hbar=c=1$ are used).
It determines the physical regime of the interaction. The motion and radiation of a particle characterized by low $\chi$ value are described by the classical electrodynamics \cite{Jackson_b_1975}.
In the opposite case, however, the recoil caused by a single photon emission is significant, introducing quantum effect to the dynamics. This phenomenon was recently reported \cite{Cole_2018,Poder_2018} and is expected to become dominant in the experiments for  next generation laser facilities \cite{ELI,XCELS}.

The appropriate framework to address laser-particle interaction corresponding to high $\xi,\chi$ values is the strong field QED \cite{RMP_2012}. Usually the Furry picture is applied \cite{Furry_1951}, describing the strong background field as a classical coherent field, and using the Dirac equation solutions in this field as the basis to the QED perturbation theory.
% strong field In the realm of this theory, the electromagnetic field is considered as a classical potential \cite{Furry_1951}, appearing in the Dirac equation. The corresponding solutions are the basis to the perturbative theory, rather than free particle wave functions.
Unfortunately, the Dirac equation cannot be analytically solved in the presence of general background fields. Exact solutions exist only for few simplified configurations  \cite{Volkov_1935,Klepikov_1954,Sokolov_1986,Redmond_1965,Nikishov_1971}.
Most of the
%recent
research in the context of strong field QED is based on the Volkov wave function \cite{Volkov_1935,Ritus_1985}, describing a particle in the presence of a plane wave field. While fully adequate to account for the interaction of electrons with a  single laser beam,
%a beam with a single laser,
the Volkov solution cannot be used in the presence of a tightly focused beam,
multiple beam configuration, or plasma self-generated fields.
The common practice in these cases is to assume that in the presence of strong fields
the radiation is determined by the local value of the external field, due to smallness of the radiation formation length in ultrastrong fields with respect to the laser wavelength. This approach, known as the Local Constant Field Approximation (LCFA), is the basis of the PIC-QED codes \cite{Elkina_2011,Gonoskov_2015}, which are the main tool to explore laser-plasma interaction in the high-intensities regime. Notwithstanding of the success of LCFA approach, several recent works \cite{DiPiazza_2018,DiPiazza_2019,raicher_2019,Lv_2020} pointed out its limitations and possible violations
%of the LCFA
under certain conditions.

Thus, theoretical investigation of QED processes in the presence of complex field configurations
%beyond LCFA
requires alternative approaches beyond LCFA. One way is provided by
% such as
the semiclassical operator techniques introduced by Schwinger \cite{Schwinger_1954}, and further significantly developed by Baier and Katkov \cite{Katkov_1968, Baier_b_1994, Landau_4}.  The Baier-Katkov method is applicable when the electron dynamics in the background fields is quasiclassical and it allows for the evaluation of quantum radiation with a large quantum recoil at a photon emission. In this approach,  the photon emission probability amplitude is expressed in terms of the classical trajectory, which effectively means using the Wentzel-Kramers-Brillouin (WKB) approximation in the leading order for the electron wave function in background field.
% rather than the wave function.
The Baier-Katkov method thus enables the evaluation of quantum radiation in scenarios where the electron motion is quasiclassical and the electron's classical trajectory is known.
The method's deficiency is that it is limited to tree-level strong-field QED processes.
%The method's deficiency is that it does not allow for the systematic improvement of the WKB approximation for treating electron dynamics.
%in scenarios where  and the electron's classical trajectory is known. In this approach, they expressed the photon emission probability (including quantum recoil effects) in terms of the classical trajectory rather than the wave function. The Baier-Katkov method thus enables the evaluation of quantum radiation in scenarios where the electron motion is quasiclassical and the electron's classical trajectory is known.
%This approach was used for several strong field configurations \cite{}. This approach, however, cannot provide the wavefunction of the particle under consideration. However, this approach accounts only for quantum effects during the radiation and neglects quantum effects on the electron's dynamics.

To achieve a more comprehensive quantum description of QED processes in strong fields, one can explicitly derive the electron wave function in the background field within the WKB approximation and use it in the Furry picture for the calculation of amplitudes of QED processes. Recently, this approach has been applied to investigate strong-field QED processes in tightly focused laser beams \cite{DiPiazza_2015, DiPiazza_2016, DiPiazza_2017}.

 %employ the WKB approximation %\cite{Popov_1997, Gersten_1975, Mocken_2010} to obtain the wave function in the background field. Then, applying this wave function in the Furry picture framework to analyze strong-field QED processes. Recently, this approach has been applied to investigate strong-field QED processes in tightly focused laser beams \cite{DiPiazza_2015, DiPiazza_2016, DiPiazza_2017}.

Tightly focusing laser beam is used in experiments to enhance the field strength for a given laser beam energy. However, the same goal can be achieved in multi-beam configurations. For instance, recently it has been proposed to create the so-called dipole wave \cite{Bulanov_2010_a,Gonoskov_2013,Gonoskov_2014} using combination of multiple laser beams.
%Beyond tightly focused laser beam configurations, another approach to enhance the field strength for a given laser beam energy is the use of multi-beam configurations or the so-called dipole wave \cite{Bulanov_2010_a, Golla_2012, Gonoskov_2012, Bashinov_2013, Bashinov_2019, Magnusson_2019}.
The simplest example of a multi-beam configuration is the setup of counterpropagating waves (CPW), which provides a compelling platform for studying QED effects \cite{Kirk_2009, Bulanov_2010, Gonoskov_2014, Gong_2017, Grismayer_2017} and is particularly conducive to the generation of QED cascades \cite{Grismayer_2016, Jirka_2016}.

The dynamics of a particle in CPW field configuration strongly depends on the ratio of the laser frequencies observed in the electron's average rest frame. When this ratio approaches unity, the system enters a resonant regime, leading to phenomena such as the Kapitza-Dirac effect \cite{Kapitza_1933, Batelaan_2007, Ahrens_2012, Mueller_2017} and stimulated Compton emission \cite{Friedman_1988, Pantell_1968, Fedorov_1981,Avetissian_b_2016}. Conversely, when the frequency ratio is significantly greater than one, the electrons dynamics is nonresonant and the emitted radiation has spontaneous character rather than stimulated one, exhibiting entirely different features, as explored in various studies \cite{King_2016,Hu_2015,Lv_2020b}. Note that the CPW configuration can induce dynamical trapping, which is highly sensitive to the nature of the emission process \cite{Gonoskov_2014,Kirk_2016}.

In this paper an approximate solution to the quantum equation of motion for both scalar particles and fermions in the CPW configuration is derived using the WKB technique. Our starting point is the approximate analytical solution of the classical equations of motion obtained by us in Ref.~ \cite{Lv_2020b}. For the quantum problem, we discuss the same setup, when in the ultrarelativistic electron rest frame the counterpropagating laser wave frequency significantly exceeds that of the copropagating one, and apply the same additional approximations as in the classical case.
%trajectory recently obtained by the authors \cite{Lv_2020b}.
In virtue of the classical solution, the classical action is found, which further facilitates the construction of the WKB solution. For the derivation of the time-dependent spinorial part of the wave function an iterative technique is applied. With the derived wave function,  several quantum characteristics of the dynamics are analyzed in detail.
%Then, a WKB solution is constructed and  analysis of its validity condition is carried out.

The paper is organized as follows. Section.~\ref{sec:class} describes the classical theory:
\ref{subsec:traj} summarizes the results derived in Ref.~\cite{Lv_2020b} regarding the electron's classical trajectory in CPW and \ref{subsec:action} derives the classical action. The WKB wave function as well as its validity criteria appear in Sec.~\ref{sec:WKBwave}. In Sec.~\ref{sec:num}, the spin dynamics resulting from the novel wave function is demonstrated. Sec.~\ref{sec:con} concludes the main findings the work and gives some outlook for the application of the wave function in future work.

\section{Classical theory}
\label{sec:class}
\subsection{Classical trajectory}

\label{subsec:traj}
In this subsection the mathematical formulation of the CPW problem is introduced. The approximate solution of the classical equation of motion for a relativistic electron in the CPW setup and its validity condition, derived in
%a
our previous work \cite{Lv_2020b}, are summarized. The classical equation of motion for a particle in the presence of an electromagnetic field reads
\begin{equation}
  \label{eq:av EOM}
  \frac{d P^{\mu}}{d \tau} = \frac{e}{m} F^{\mu \nu} P_{\nu} ,
\end{equation}
where $\tau$ is the proper time, $P_{\mu}$ is the particle's 4-momentum and $F_{\mu \nu} \equiv \partial_{\mu} A_{\nu}- \partial_{\nu} A_{\mu} $ is the electromagnetic field tensor. The vector potential corresponding to the circularly polarized CPW configuration is $A^{\mu}=A^{\mu}_1+A^{\mu}_2$ where
\begin{equation}
  \label{eq:av A_def1}
  A^{\mu}_1 \left( \phi_1 \right) \equiv a_1 \left[ \cos (\phi_1) e^{\mu}_x + \sin (\phi_1) e^{\mu}_y \right] ,
\end{equation}
\begin{equation}
  \label{eq:av A_def2}
  A^{\mu}_2 \left( \phi_2 \right) \equiv a_2 \left[ \cos (\phi_1) e^{\mu}_x + \sin (\phi_1) e^{\mu}_y \right] ,
\end{equation}
where $\phi_1=k_1 \cdot x$ and $\phi_2=k_2 \cdot x$. The wave vectors read
\begin{equation}
  \label{eq:av k_def}
  k_1 = (\omega,0,0,\omega), \quad \quad k_2 = (\omega,0,0,-\omega).	
\end{equation}
The scalars $a_1,a_2$ are the fields' amplitudes. In the following we use their normalized values $\xi_1=ea_1/m$, and $\xi_2=ea_2/m$, and $e_x = (0,1,0,0), e_y = (0,0,1,0)$ are unit vectors.

The key assumption, allowing
%one
us to perform the integration over the proper time to solve Eq.~\eqref{eq:av EOM}, is that the particle propagates with relativistic velocity  at a small angle with respect to the CPW axis.
%In other words,
In mathematical terms, the following conditions should be fulfilled
\begin{eqnarray}
  \frac{m \xi_1 p_x \omega}{(k_1 \cdot \bar{P})\bar{\varepsilon}} & \ll & 1 ,  \label{eq:av condP1} \\
  \frac{m \xi_2 p_x \omega}{(k_2 \cdot \bar{P})\bar{\varepsilon}} &\ll& 1 . \label{eq:av condP2}\\
  \frac{ 2m^2 \xi_1 \xi_2 \omega}{ [\Delta k \cdot \bar{P}]\bar{\varepsilon}} & \ll & 1 , \label{eq:av condP3}
\end{eqnarray}
where $\Delta k=k_1-k_2$.
It should be noticed that for the sake of simplicity, these conditions are expressed in terms of the quantities in the frame of reference where the counterpropagating waves have the same frequency, rather than in a Lorentz invariant form.
The above expressions depend on the time-averaged 4-momentum $\bar{P}_{\mu}=\left( {\bar{\varepsilon},p_x,0,\bar{P}_z} \right)$, which is specified below.
Due to the azimuthal symmetry we assume, without loss of generality, that the transverse momentum lies in the $x$ axis. %$p_{\bot}=p_x$.

In this setup, as is shown in \cite{Lv_2020b}  the following relation is approximately fulfilled
%Consequently, it is shown in \cite{Lv_2020b} that the following relation is obtain
\begin{equation}
  \label{eq:av key_approx}
  \int d \tau f' \left( \phi_1 \right) \approx \frac{m}{k_1 \cdot P} f(\phi_1),
\end{equation}
as well as an analogous
%identity
one with $1 \rightarrow 2$. The symbol $f(\phi_1)$ stands for either $\sin \phi_1$ or $\cos \phi_1$.
Using (\ref{eq:av key_approx}), the Lorentz equation can be integrated and the momentum takes the form
\begin{multline}
  \label{eq:av final_LI}
  P_{\mu}(\tau)= \bar{P}_{\mu}-e\left[ A^{\mu}_1(\phi_1)+A^{\mu}_2(\phi_2) \right] \\
                +k^{\mu}_1 \left[ \frac{ep \cdot A_1(\phi_1)}{k_1 \cdot \bar{P}} - \frac{ A_1(\phi_2) \cdot A_2(\phi_2) }{ \Delta k \cdot \bar{P}}  \right] \\
                +k^{\mu}_2 \left[ \frac{ep \cdot A_2(\phi_2)}{k_2 \cdot \bar{P}} + \frac{ A_1(\phi_2) \cdot A_2(\phi_2) }{ \Delta k \cdot \bar{P}}  \right] .
\end{multline}
The phases are given by
\begin{eqnarray}
  \phi_1(\tau) &=&  \frac{k_1 \cdot \bar{P}}{m} \tau , \label{eq:av phi1F}\\
  \phi_2(\tau)   &=& \frac{k_2 \cdot \bar{P}}{m} \tau  + \frac{2 p_x m \xi_1 \omega^2}{(k_1 \cdot \bar{P})^2} \sin \phi_1. \label{eq:av phi2F}
\end{eqnarray}
Without lose of generality, we have chosen the electron to be copropagating with the laser beam $\xi_1$.
% here. The trajectory is related to the momentum according to
We derive the trajectory via
\begin{equation}
  \label{eq:av traj_i}
  x_{\mu}(\tau)= \int d \tau \frac{P_{\mu}(\tau)}{m} ,
\end{equation}
which yields for its components
%and its components are correspondingly given by
\begin{eqnarray}
  x(\tau) &=&  \left(\frac{p_x}{m} \tau+ \frac{m \xi_1}{k_1 \cdot \bar{P}} \sin \phi_1 + \frac{m \xi_2}{k_2 \cdot \bar{P}} \sin \phi_2 \right) \,,\label{eq:av final_x}\\
  y(\tau) &=& - \left( \frac{m \xi_1}{k_1 \cdot \bar{P}} \cos \phi_1 + \frac{m \xi_2}{k_2 \cdot \bar{P}} \cos \phi_2 \right) \,. \label{eq:av final_y}\\
%\end{eqnarray} and \begin{multline}
  \label{eq:av final_z}
  z(\tau) &=& \frac{\bar{P}_z}{m} \tau +\frac{ 2m^2 \xi_1 \xi_2 \omega }{ \left(\Delta k \cdot \bar{P}\right)^2} \sin \left( \phi_1-\phi_2 \right) \\
            &+& p_x \omega \left[ \frac{m \xi_1}{(k_1 \cdot \bar{P})^2} \sin \phi_1 - \frac{m \xi_2}{(k_2 \cdot \bar{P})^2} \sin \phi_2 \right].\nonumber
\end{eqnarray}
%\end{multline} Finally, one should find the relation between the asymptotic momentum and the average one. It was shown in \cite{Lv_2020b} to depend on the turn-on process of the beams.
The relation between the asymptotic momentum and the average one, as shown in Ref.~\cite{Lv_2020b},  depends on the turn-on process of the beams. If the beams are turned on separately,
%one can write down
an analytical expression exists for the average momentum. In case beam 1 is turned on first, we have
\begin{equation}
  \label{eq:av barPz1}
  \bar{P}^{\mu} =p^{\mu}+  \frac{m^2 \xi_1^2}{2( k_1 \cdot p)} k_1^{\mu}+\frac{m^2 \xi_2^2}{2 [k_2 \cdot \bar{P}^{(1)}]} k_2^{\mu} ,
\end{equation}
where
\begin{equation}
  \bar{P}^{(1)}_{\mu} =p_{\mu}+ \frac{m^2 \xi_1^2}{2( k_1 \cdot p) } k_{1 \mu}.
\end{equation}
In case the second beam is turned on first, analogous derivation leads to
\begin{equation}
  \label{eq:av barPz2}
  \bar{P}^{\mu} =p^{\mu}+  \frac{m^2 \xi_1^2}{2[k_1 \cdot \bar{P}^{(2)}]} k_1^{\mu}+\frac{m^2 \xi_2^2}{2 (k_2 \cdot p)} k_2^{\mu},
\end{equation}
where
\begin{equation}
  \bar{P}^{(2)}_{\mu} =p_{\mu}+ \frac{m^2 \xi_1^2}{2 (k_2 \cdot p )} k_{2 \mu}.
\end{equation}
It would prove useful to write down explicitly the Lorentz invariant quantities $k_1 \cdot \bar{P},k_2 \cdot \bar{P}$ as
\begin{equation}
  \label{eq:av om12}
  k_1 \cdot \bar{P}= \omega_1 \bar{\varepsilon}, \quad  k_2 \cdot \bar{P}= \omega_2 \bar{\varepsilon}.
\end{equation}
The frequencies $\omega_1,\omega_2$ are related to the average velocity along $z$ axis $\bar{v}_z=\bar{P}_z/\bar{\varepsilon}$ by $\omega_1 \equiv \omega (1-\bar{v}_z)\approx \frac{\omega}{2} \left(\frac{m_*}{\bar{\varepsilon}}\right)^2$ and $ \omega_2 \equiv \omega (1+\bar{v}_z) \approx 2 \omega$.
The effective mass is defined as $m_* \equiv \sqrt{\bar{P}^2}$.
In our setup, in Lab-frame the counterpropagating laser waves have the same frequency $\omega$, and the electron (copropagating with $\xi_1$) is ultrarelativistic, accordingly $\omega_2/\omega_1 \approx 4\gamma_*^2\gg 1$, with $\gamma_*=\bar{\varepsilon}/m_*$. Using Eq.~(\ref{eq:av om12}), the restrictions imposed by 
the conditions (\ref{eq:av condP1})-(\ref{eq:av condP2}) on the average electron propagation angle with respect to the laser beam axis ($\theta=p_x / \bar{\varepsilon} $) may be expressed by
$m\xi_1\bar{\varepsilon} \theta / m_*^2 \ll 1$ and $m\xi_2\theta/ \bar{\varepsilon}  \ll 1$. 
%$\xi_1\bar{\gamma} \theta\ll 1$, and $\xi_2\theta/ \bar{\gamma}  \ll 1$. 

%\section{The WKB wave function}\label{sec:WKBwave} The obtained analytical solution of classical equations of motion in counterpropagating laser fields can be employed to construct the wave function of the electron in the framework of the WKB approximation. As a first step we derive the classical action in this setup.

\subsection{The classical action}
\label{subsec:action}
The classical action is defined \cite{Landau_2} as
%may be found by integrating the following equation \cite{Landau_2}
\begin{equation}
  \label{eq:av action_tau1}
  S=-\int \left(m+\frac{e}{m} P \cdot A \right)d \tau .
\end{equation}
Employing the final momentum obtained above in Eq.~(\ref{eq:av final_LI}), as well as the vector potential of the fields defined in Eqs.~\eqref{eq:av A_def1}-\eqref{eq:av A_def2} and the approximation Eq.~\eqref{eq:av key_approx}, the integration is carried out
\begin{multline}
  \label{eq:av action_tau2}
  S( \tau) = -\frac{m_*^2}{m} \tau - \frac{2m^2 \xi_1 \xi_2}{\Delta k \cdot \bar{P}} \sin (\phi_1-\phi_2) \\
            -p_x \left[ \frac{m \xi_1}{k_1 \cdot \bar{P}} \sin \phi_1 + \frac{m \xi_2}{k_2 \cdot \bar{P}} \sin \phi_2  \right] ,
\end{multline}
where $\phi_1(\tau), \phi_2(\tau)$ are given in Eqs.~(\ref{eq:av phi1F},\ref{eq:av phi2F}).
Our final goal is to use the action to construct the WKB wave function. To this end we need to express it in terms of $x_{\mu}$.
%For the sake of this purpose, we wish to express it in terms of $x_{\mu}$ rather than $\tau$. The first step is to replace the first term in (\ref{eq:av action_tau2}).
For this purpose, we multiply  the trajectory obtained in Sec. II by $\bar{P}$:
%, one may show
\begin{equation}
  \label{eq:av action_tau3}
  \bar{P} \cdot x(\tau) = \frac{m_*^2}{m} \tau  + \frac{m^2 \xi_1 \xi_2}{\Delta k \cdot \bar{P}} \sin (\phi_1-\phi_2).
\end{equation}
Owing to this relation, the action is expressed as
%may  written as
\begin{multline}
  \label{eq:av action_tau4}
  S( \tau) = -\bar{P} \cdot x(\tau) - \frac{m^2 \xi_1 \xi_2}{\Delta k \cdot \bar{P}} \sin (\phi_1-\phi_2) \\
            -p_x \left[ \frac{m \xi_1}{k_1 \cdot \bar{P}} \sin \phi_1 + \frac{m \xi_2}{k_2 \cdot \bar{P}} \sin \phi_2  \right] .
\end{multline}
Next, we recall that $\phi_1,\phi_2$ depend on the proper time through $\phi_1=k_1 \cdot x(\tau), \phi_2=k_2 \cdot x(\tau)$.
Hence, in order to obtain $S(x) $ one should simply replace $x_{\mu}(\tau)$ by $x_{\mu}$, leading to
\begin{multline}
  \label{eq:av action_f}
  S(x) = -\bar{P} \cdot x - \frac{m \xi_1 \xi_2}{\Delta k \cdot \bar{P}} \sin \left( \Delta k \cdot x \right) \\
        -p_x \left[ \frac{m \xi_1}{k_1 \cdot \bar{P}} \sin (k_1 \cdot x) + \frac{m \xi_2}{k_2 \cdot \bar{P}} \sin (k_2 \cdot x)  \right] .
\end{multline}
One can explicitly verify this action by substituting it into the Hamilton-Jacoby relation \cite{Landau_2}
\begin{equation}
  \label{eq:av HJE}
  P_{\mu}= - \partial_{\mu} S-eA_{\mu} ,
\end{equation}
which yields indeed the 4-momentum obtained in the previous subsection.
In the case when one of the fields vanishes, Eq.~(\ref{eq:av action_f}) recovers the familiar electron action in a plane wave laser field \cite{Volkov_1935}.\\

\section{The WKB wave function}
\label{sec:WKBwave} 

The obtained analytical solution of the classical equations of motion in counterpropagating laser fields can be employed to construct the wave function of the electron in the framework of the WKB approximation. 

\subsection{Solution of Klein-Gordon equation}

In the following, we derive the relativistic electron wave function in WKB approximation, beginning with the spinless case.
%we utilize the classical action to derive an approximate solution for the quantum dynamics. We begin with the spinless case and later extend the discussion to include fermions.
The wave function of a scalar particle in the CPW field configuration satisfies the Klein-Gordon equation:
\begin{equation}
  \label{eq:av KG1}
  \left[\left( i\partial_{\mu}-eA_{\mu} \right)^2 - m^2 \right] \psi  =0 .
\end{equation}
We employ the WKB Ansatz \cite{Akhiezer_1965}, expressing the solution as a series in $\hbar$:
\begin{equation}
  \label{eq:av WKB_ansatz}
  \psi= e^{iS/ \hbar} \left( U_0+U_1 \hbar+...\right).
\end{equation}
As the WKB method is a perturbative approach with respect to $\hbar$, we explicitly incorporate $\hbar$ into the equations presented in this section for clarity.
% and precision.
%Since the WKB is a perturbation theory with respect to $\hbar$, we have explicitly write it out in the equations appearing in this section.
% Notice that for the sake of this derivation, $\hbar$ is written explicitly, whereas in the rest of the paper natural units $\hbar=1$ are used.
%Applying perturbation theory with respect to $\hbar$,
We obtain in
%The WKB Ansatz (\ref{eq:av WKB_ansatz}) is substituted into the equation of motion (\ref{eq:av KG1}), and the various powers of $\hbar$ are considered separately.
the leading order ($\sim \hbar^0$)
%(containing no $\hbar$)
the classical Hamilton-Jacobi equation \cite{Landau_2}:
%yields
\begin{equation}
  \label{eq:av zerothM}
  \left(  \partial_{\mu} S+eA_{\mu} \right)^2-m^2=0 ,
\end{equation}
which has an approximate solution  given by Eq.~(\ref{eq:av action_f}).
%This equation is an alternative representation of the Hamilton-Jacobi equation (\ref{eq:av HJE}), since $P^2=m^2$, as is explicitly shown in \cite{Landau_2}.
The equation for the first order correction ($\sim \hbar$) reads
%of the wave function in $\hbar$ reads
\begin{equation}
  \label{eq:av firstM}
  \left[ \partial^{\mu} \left(  \partial_{\mu} S+eA_{\mu} \right) \right]U_0+2 \left(  \partial_{\mu} S+eA_{\mu} \right)  \partial^{\mu} U_0 =0 .
\end{equation}
%In the following the higher orders in $\hbar$ are omitted.
Using Eq.~(\ref{eq:av HJE}), the equation for $U_0$ takes the form
\begin{equation}
  \label{eq:av R1}
  \frac{1}{2} \partial_{\mu} P^{\mu}  U_0 +P^{\mu} \partial_{\mu} U_0=0.
\end{equation}
This partial differential equation of first order is solved using the characteristics method \cite{Courant_1998}. Accordingly, the coordinates $x_{\mu}$ are given in terms of the characteristic variable $s$
\begin{equation}
\frac{d x_{\mu}}{ds}=P_{\mu},
 \label{eq:av WKB1}
\end{equation}
Comparing with (\ref{eq:av traj_i}) one observes that
the characteristic variable is proportional to the proper time $s=\tau/m$. Then, Eq.~(\ref{eq:av R1}) reads
%With respect to $U_0$, this method yields the following ODE
\begin{equation}
\frac{dU_0}{d \tau}=-
\frac{1}{2m}   \partial_{\mu} P^{\mu} U_0,
\end{equation}
with the solution
%is given by
\begin{equation}
U_0= \exp \left[
- \int \frac{d \tau}{2m}  \partial_{\mu} P^{\mu}\right] .
\end{equation}
Let us explicitly evaluate the integrand, employing the 4-momentum of Eq.~(\ref{eq:av final_LI})
\begin{equation}
\partial_{\mu} P^{\mu}=-\frac{2m^2 \xi_1 \xi_2 (k_1 \cdot k_2)}{\Delta k  \cdot \bar{P}} \sin \left( \phi_1 - \phi_2 \right).
 \label{eq:av doP}
 \end{equation}
Integrating over $\tau$  we obtain
\begin{equation}
U_0=\exp \Biggl[  \frac{m^2 \xi_1 \xi_2 (k_1 \cdot k_2)}{[\Delta k \cdot \bar{P}]^2} \cos \left( \phi_1 - \phi_2 \right)
 \Biggr].
 \label{eq:av final_R}
\end{equation}
Since we wish to express the wave function as a function of $x_{\mu}$ rather than $\tau$, the phases $\phi_1,\phi_2$ are replaced by $k_1 \cdot x, k_2 \cdot x$, respectively, as discussed in the previous section.
Furthermore, one notices that according  to our key assumption (\ref{eq:av condP3}),
%(\ref{eq:av key_approx}),
the exponent's argument is much smaller than 1, and one may therefore Taylor expand it as
\begin{equation}
U_0 \approx 1+  \frac{m^2 \xi_1 \xi_2 (k_1 \cdot k_2)}{[\Delta k \cdot \bar{P}]^2} \cos \left[ \Delta k \cdot x \right].
 \label{eq:av final_R2}
\end{equation}
Consequently, the first order WKB solution is obtained
%up to the normalization factor. The corresponding wave function takes the form
\begin{equation}
\psi =\frac{C_N}{\sqrt{2 \varepsilon_0 V}} U_0 e^{iS},
 \label{eq:av final_psiKG}
\end{equation}
where $V $ is the interaction volume and $1/\sqrt{2 \varepsilon_0 V}$ is the free particle normalization factor.
%Hence,
The  coefficient $C_N$ is determined according to the charge normalization
%equals . This coefficient is determined according to the charge normalization condition
\begin{equation}
\int{d^3x} j_0=1,
 \label{eq:av normalization}
\end{equation}
with the Klein-Gordon 4-current
%is related to the wave function by
\begin{equation}
j _{\mu} =   \psi^*  \left(i\partial_{\mu}-eA_{\mu} \right) \psi -\psi  \left(i\partial_{\mu}+eA_{\mu} \right) \psi^*  ,
 \label{eq:av current}
\end{equation}
and $C_N=1$ at vanishing electromagnetic fields. With Eqs.~(\ref{eq:av final_psiKG}) and (\ref{eq:av current})
one obtains 
\begin{equation}
j_{\mu} =\left(C_NU_0 \right)^2 \frac{P_{\mu}}{\varepsilon_0 V}.
 \label{eq:av current2}
\end{equation}
%Plugging the charge density $j_0$ to (\ref{eq:av normalization}) and writing $R_0,\varepsilon$ explicitly we find
Then Eq.~(\ref{eq:av normalization}) can be rewritten as
\begin{multline}
 \frac{C^2_N}{\varepsilon_0 V} \int{d^3x} \left[
1+  \frac{m^2 \xi_1 \xi_2 (k_1 \cdot k_2)}{[\Delta k \cdot \bar{P}]^2} \cos \left[ \Delta k \cdot x \right] \right]^2 \times\\
\left[\bar{\varepsilon}+
\omega \left(\frac{p_x m \xi_1 \cos(k_1 \cdot x)}{k_1 \cdot \bar{P}}
+\frac{p_x m \xi_2 \cos(k_2 \cdot x)}{k_2 \cdot \bar{P}}
\right) \right]
 =1.
\end{multline}
We notice that the oscillatory parts of both $U_0$ and $\varepsilon$ are small as compared to their average value. When multiplying both expressions, the second order terms are discarded. The first order terms are either sine or cosine functions with respect to the spatial coordinates, and therefore their contribution to the integration vanishes. Thus,
%the integral yields a constant and
the normalization coefficient reads
\begin{equation}
C_N=\sqrt{\frac{\varepsilon_0}{\bar{\varepsilon}}}.
 \label{eq:av normcoefKG}
\end{equation}
Finally, it should be noted that in the case when one of the laser waves vanishes, $U_0=1$, the action $S$ reduces to its plane wave value, and the Volkov solution \cite{Landau_4} is recovered.\\
%verify that in case one the waves vanishes the Volkov solution is recovered. From (\ref{eq:av final_R2}) one can instantly see that in this case $R_0=1$. Moreover, the action $S$ recovers its plane wave value, as shown above. Therefore, our expression (\ref{eq:av final_psiKG}) indeed reduces to the familiar Volkov solution \cite{Landau_4}.

Let us discuss now the validity domain of the solution.
One may verify that the WKB approximation presented above is equivalent to neglecting the term $\hbar^2 \partial^2 U_0$ after substituting the Ansatz $\psi= U_0 e^{iS/ \hbar} $ to the equation of motion (\ref{eq:av KG1}). Hence, in order to estimate its validity regime, let us compare the neglected high-order term
%residue
\begin{equation}
U^{(2)} \sim \partial^2 U_0 \sim \frac{m^2 \xi_1 \xi_2 \left( k_1 \cdot k_2 \right) \Delta k^2}{\left( \Delta k \cdot P \right)^2} \sim \frac{m \xi_1 \xi_2 \omega^2}{\bar{\varepsilon}^2},
\end{equation}
with the first-order one
%correction terms
\begin{equation}
U^{(1)} \sim  \left( \partial_{\mu} P^{\mu} \right) U_0 \sim \frac{m^2 \xi_1 \xi_2 \left( k_1 \cdot k_2 \right) }{\Delta k \cdot P} \sim \frac{m \xi_1 \xi_2 \omega}{\bar{\varepsilon}}
\end{equation}
where $U_0 \sim 1$.
%Namely,
Consequently, the WKB solution is
%expected to be
valid as long as the following condition is met
\begin{equation}
  \label{eq:av WKB_cond}
  \frac{ U^{(2)}}{ U^{(1)}} \sim \frac{\omega}{\bar{\varepsilon}}\ll 1,
\end{equation}
which coincides with the Baier-Katkov semiclassical condition \cite{Baier_b_1994}.

\subsection{Solution of Dirac equation}

We now extend our calculations to the case of a fermion in CPW configuration, focusing on solving the Dirac equation
%Now let us extend our calculation to the case of a fermion in a counterpropagating laser fields, namely solving Dirac equation
% The corresponding quantum equation of motion is the Dirac equation
\begin{equation}
  \label{eq:av Dirac0}
  \left( i \slashed{\partial}-e \slashed{A}-m \right) \Psi  =0,
\end{equation}
where the slash symbol stands for $\slashed{u} \equiv \gamma \cdot u$ for a general four-vector $u_{\mu}$. $\gamma_{\mu}$ are the Dirac matrices and the wave function $\Psi$ is a four-component bispinor.
%\cite{Landau_4}.  For the sake of this derivation it is convenient to transform into a second-order equation,
It is convenient to use the quadratic form of the Dirac equation, resembling the Klein-Gordon one discussed above. Applying the operator $\left( i \slashed{\partial}-e \slashed{A}+m \right)$ to Eq.~\eqref{eq:av Dirac0} one obtains \cite{Landau_4}
\begin{equation}
  \label{eq:av Dirac1}
  \left[\left( i\partial_{\mu}-eA_{\mu} \right)^2 - m^2 -\frac{ie}{2} \sigma_{\mu \nu} F^{\mu \nu} \right] \Psi =0,
\end{equation}
where the $\sigma_{\mu \nu} \equiv \frac{1}{2} (\gamma_{\mu} \gamma_{\nu} - \gamma_{\nu} \gamma_{\mu})$.
The WKB Ansatz reads $\Psi= {\cal U}_0 e^{iS/ \hbar}$ where ${\cal U}_0$ is now a spinor.
%(\ref{eq:av WKB_ansatz}),
% We repeat the same procedure as in the previous subsection and observe that
While the equation for $S$ remains the same as in the Klein-Gordon case, namely Eq.~(\ref{eq:av zerothM}),
%satisfied by $S$ remains the same, namely (\ref{eq:av zerothM}).
the equation for the first order in $\hbar$ contains an additional term as compared to the scalar version (\ref{eq:av R1})
\begin{multline}
  \label{eq:av firstMD}
  \left[ \partial^{\mu} \left(  \partial_{\mu} S+eA_{\mu} \right) \right]{\cal U}_0+2 \left(  \partial_{\mu} S+eA_{\mu} \right)  \partial^{\mu} {\cal U}_0 \\+\frac{e}{2} \sigma_{\mu \nu} F^{\mu \nu} {\cal U}_0=0 .
\end{multline}
Using Eq.~\eqref{eq:av A_def1} and \eqref{eq:av A_def2} and the definition of the field tensor $F_{\mu \nu}$, the spinor part takes the form
\begin{equation}
  \label{eq:av Dirac1}
  \frac{e}{2} \sigma_{\mu \nu} F^{\mu \nu}= e (\slashed{k}_1 \slashed{A}'_1+\slashed{k}_2 \slashed{A}'_2),
\end{equation}
where the symbol ' designates derivative with respect to $\phi_1$ or $\phi_2$, respectively. Employing as before the characteristics method one arrives at
%an ODE for ${\cal R}_0$ follows
\begin{equation}
  \label{eq:av eq_spin}
  \frac{d{\cal U}_0}{d \tau} = -\frac{1}{2m} \left[\partial_{\mu} P^{\mu} - e (\slashed{k}_1 \slashed{A}'_1+\slashed{k}_2 \slashed{A}'_2) \right] {\cal U}_0.
\end{equation}
We substitute ${\cal U}_0= U_0\left( \tau \right) \Omega \left( \tau \right)$ where $U_0\left( \tau \right)$ was derived in the previous section (\ref{eq:av final_R}) and obtain
\begin{equation}
\frac{d \Omega}{d \tau}=
\Lambda \left(\tau \right) \Omega, \quad \quad \Lambda \left(\tau \right) \equiv
    \frac{e (\slashed{k}_1 \slashed{A}'_1+\slashed{k}_2 \slashed{A}'_2)}{2m} .
\label{eq:av Om_eq}
\end{equation}
As opposed to the scalar case, the equation above does not admit an exact solution, since the matrices $\slashed{k}_1 \slashed{A}_1 (\tau), \slashed{k}_2 \slashed{A}_2 (\tau)$ do not commute at different times.
A possible analytical approximation to this class of mathematical problems is the Magnus expansion \cite{magnus1954exponential,blanes2009magnus}. However, in this particular case a second order Magnus expansion was not satisfactory and high order terms turned out to be very cumbersome. Therefore, we put forward an alternative iterative approach.
% expansion is suggested.

We notice that, had Eq. (\ref{eq:av Om_eq}) been a scalar one, its solution would simply be $\Omega(\tau) = e^{\int{\Lambda(\tau) d \tau}}$. Let us examine the series corresponding to the Taylor expansion of the exponent
\begin{equation}
\Omega \left(\tau \right) \equiv
\sum_{n=0}^{N}{\Omega_n \left( \tau \right)},
\label{eq:av recursive1}
\end{equation}
with $\Omega_n = \frac{1}{n!} \left(\int{\Lambda(\tau) d \tau}  \right)^n$ and $N \rightarrow \infty$. One may readily verify that successive terms satisfy the following recursive relation
\begin{equation}
\frac{d \Omega_n}{d \tau}=\Lambda \left( \tau \right) \Omega_{n-1}.
\label{eq:av recursive2}
\end{equation}
Inspired by this fact, we seek a solution for the spinor equation (\ref{eq:av Om_eq}) in the form of Eqs.~(\ref{eq:av recursive1})-(\ref{eq:av recursive2}).
By direct substitution of the Ansatz Eq.~\eqref{eq:av recursive1} into Eq.~\eqref{eq:av Om_eq} and assuming Eq.~\eqref{eq:av recursive2} is fulfilled, one obtains that the equation is satisfied up to the residual term
\begin{equation}
  \label{eq:av residue}
  {\cal R} \equiv \frac{d \Omega}{d \tau}-\Lambda \Omega = \Lambda \Omega_{n+1}.
\end{equation}
In the following, Eq. (\ref{eq:av recursive2}) is solved, yielding a general expression for $\Omega_n$. We explicitly demonstrate that each term is smaller then the previous one, so the total sum converges to the equation's solution.

The first step to be taken is to specify the zeroth-order term. We set it to be $\Omega_0 =u_{p_0,\sigma_0}$ where $u_{p_0,\sigma_0}$ is the field free particle bispinor with the normalization being $u_{p,\sigma}^{\dagger}u_{p',\sigma'}=\delta_{p,p'}\delta_{\sigma,\sigma'}$.
This choice ensures that our solution satisfies the initial conditions at $t \rightarrow -\infty$. Thus, the redundant unphysical solution associated with the second order Dirac equation is excluded \cite{Landau_4}.
The first-order term straightforwardly reads
\begin{equation}
\Omega_1(\tau)= \frac{e}{2} \left( \frac{\slashed{k}_1 \slashed{A}_1}{k_1 \cdot \bar{P}}+\frac{\slashed{k}_2 \slashed{A}_2}{k_2 \cdot \bar{P}} \right) u_{p_0,\sigma_0}.
\end{equation}
In order to solve for the second-order term, it proves convenient to write down explicitly the first beam vector potential, introduced in (\ref{eq:av A_def1}) as
\begin{equation}
e A^{\mu}_1 \left(\phi_1 \right) = \frac{m \xi_1}{2} \left[
\epsilon^{\mu} e^{i \phi_1} + \left( \epsilon^{\mu} \right)^* e^{-i \phi_1}
\right],
\label{eq:av A1_exp}
\end{equation}
as well as $1 \rightarrow 2$ for the second wave. The polarization 4-vector is $\epsilon_{\mu} = \left(0,1,-i,0 \right)$.
The vector potential's derivative with respect to $\phi_1$ is therefore
\begin{equation}
e A'^{\mu}_1 \left(\phi_1 \right) = i \frac{m \xi_1}{2} \left[
\epsilon^{\mu} e^{i \phi_1} - \left(\epsilon^{\mu} \right)^* e^{-i \phi_1}
\right].
\label{eq:av A1d_exp}
\end{equation}
Since $k_1 \cdot \epsilon = k_1 \cdot \epsilon^*=0$ as well as $k_1^2=k_2^2=0$, the following identities hold
\begin{equation}
\slashed{k}_1 \slashed{\epsilon} \slashed{k}_1 \slashed{\epsilon} =
\slashed{k}_1 \slashed{\epsilon^*} \slashed{k}_1 \slashed{\epsilon} =
\slashed{k}_1 \slashed{\epsilon} \slashed{k}_1 \slashed{\epsilon^*} =
\slashed{k}_1 \slashed{\epsilon^*} \slashed{k}_1 \slashed{\epsilon^*} =
0,
\end{equation}
and analogous ones for $k_2$.
Since the polarization vector satisfies $\epsilon \cdot \epsilon = \epsilon^* \cdot \epsilon^*=0$, one may verify that
\begin{equation}
  \slashed{k}_1 \slashed{\epsilon} \slashed{k}_2 \slashed{\epsilon} =
  \slashed{k}_1 \slashed{\epsilon^*} \slashed{k}_2 \slashed{\epsilon^*} = 0.
\end{equation}
The integration is carried out using the key approximation (\ref{eq:av key_approx}), as presented in subsection.~\ref{subsec:traj}.
Consequently, the second-order term is given by
\begin{multline}
\Omega_2 \left(\tau \right)=
\frac{m^2 \xi_1 \xi_2}{16 \left( \Delta k \cdot P \right)}
\Bigg[ \Bigg.
\frac{ \slashed{k}_1 \slashed{\epsilon} \slashed{k}_2 \slashed{\epsilon^*}}{k_2 \cdot P} e^{i \Delta \phi}+
\frac{ \slashed{k}_1 \slashed{\epsilon^*} \slashed{k}_2 \slashed{\epsilon}}{k_2 \cdot P} e^{-i \Delta \phi} - \\
\frac{ \slashed{k}_2 \slashed{\epsilon} \slashed{k}_1 \slashed{\epsilon^*}}{k_1 \cdot P} e^{-i \Delta \phi} -
\frac{ \slashed{k}_2 \slashed{\epsilon^*} \slashed{k}_1 \slashed{\epsilon}}{k_1 \cdot P} e^{i \Delta \phi}
\Bigg. \Bigg] u_{p_0,\sigma_0}.
 \label{eq:av Omega2}
\end{multline}
Employing (\ref{eq:av A1_exp},\ref{eq:av A1d_exp}) again in order to express the result in terms of $A_1,A_2$ one obtains
\begin{equation}
\Omega_2 \left(\tau \right)=
\frac{e^2}{4 \left( \Delta k \cdot P \right)}
\left[
\frac{ \slashed{k}_1 \slashed{A}_1 \slashed{k}_2 \slashed{A}_2}{k_2 \cdot P}
 -
\frac{ \slashed{k}_2 \slashed{A}_1 \slashed{k}_1 \slashed{A}_2}{k_1 \cdot P}
\right] u_{p_0,\sigma_0}.
 \label{eq:av Omega2f}
\end{equation}
One may repeat this procedure to obtain a general expression $\Omega_n$, for the recursive equation (\ref{eq:av recursive2})
\begin{multline}
\Omega_n \left(\tau \right) = \left(\frac{e}{2} \right)^n
\Bigg[ \Bigg.
\left(-1 \right)^{l_2} \frac{\left( \slashed{k_2} \slashed{A_2}\right)^{l_3} {\left( \slashed{k_1} \slashed{A_1} \slashed{k_2} \slashed{A_2}  \right)^{ l_1}}}{\mathcal{D}_1 \left(n \right)} + \\
\left(-1 \right)^{l_1} \frac{\left( \slashed{k_1} \slashed{A_1}\right)^{l_3} {\left( \slashed{k_2} \slashed{A_2} \slashed{k_1} \slashed{A_1}  \right)^{ l_1}}}{\mathcal{D}_2 \left(n \right)}
\Bigg. \Bigg] u_{p_0,\sigma_0},
 \label{eq:av Omega_n}
\end{multline}
where the exponentials $l_1(n) \equiv \lfloor \frac{ n} {2} \rfloor$,$l_2 (n) \equiv \lfloor \frac{ n+1} {2} \rfloor$ and $l_3 (n) \equiv \bmod(n,2)$ were introduced and the symbol $\lfloor \mu \rfloor$ stands for the integer part of the number $\mu$. The functions appearing in the denominator are defined as follows
\begin{equation}
\mathcal{D}_1 \left( n \right) \equiv \prod_{j=1}^{l_2} \left( \left[ \left(j-1 \right) \Delta k - k_2 \right] \cdot \bar{P} \right)
\prod_{j=1}^{l_1} \left( j \Delta k \cdot \bar{P} \right)
\end{equation}
and
\begin{equation}
\mathcal{D}_2 \left( n \right) \equiv \prod_{j=1}^{l_2} \left( \left[ \left(j-1 \right) \Delta k + k_1 \right] \cdot \bar{P} \right)
\prod_{j=1}^{l_1} \left( j \Delta k \cdot \bar{P} \right).
\end{equation}
Let us now identify the small parameter of the above expansion. For this purpose we examine two successive orders, $n$ and $n+1$, taking the form given in Eq.~(\ref{eq:av Omega_n}). Without a loss of generality, we assume that $n>1$ is even. The denominator for both terms scales as $(\omega \bar{\varepsilon})^n$.
As a result, the first term of the $(n+1)^{th}$ order differs by an additional factor scaling as $m\xi_2/ \bar{\varepsilon}$ from the first term of the $n^{th}$ order. Analogously, the second term acquires an additional factor scaling as $\sim  m\xi_1/ \bar{\varepsilon}$.
Next, we consider the $(n+1)^{th}$ and $(n+2)^{th}$ orders, and notice that the multiplying factors flip with respect to the previous case. Hence, the $(n+2)^{th}$ is smaller by a factor of $\xi_1 \xi_2 / \bar{\varepsilon}^2$ from the $n^{th}$ order. Namely this quantity is the small parameter corresponding to our expansion. It is always much smaller than $1$ according to the validity condition Eq.~(\ref{eq:av condP3}), so a convergence is always achieved.

As in the scalar case, in the absence of one of the beams our solution coincides with the Volkov wave function.
%Let us show that explicitly for the spinor part as well. Without loss of generality, suppose the first beam vanishes ($\xi_1=0$). Then
For instance, substituting $\xi_1=0$ the first two terms in the expansion read
\begin{equation}
\Omega= \Omega_0+\Omega_1= 1+\frac{e}{2} \left( \frac{\slashed{k}_2 \slashed{A}_2}{k_2 \cdot \bar{P}} \right).
\end{equation}
Higher-order terms in the expansion vanish due to $\slashed{k}_2 \slashed{A}_2 \slashed{k}_2 \slashed{A}_2 \propto k_2^2=0$. Namely, the expansion is truncated and recovers the spinor part of the Volkov wave function \cite{Landau_4}.
Notice that in the presence of both beams the expansion cannot be similarly truncated, as $\slashed{k}_1 \slashed{A}_1 \slashed{k}_2 \slashed{A}_2 \neq 0$.
As in the scalar case, we may now write the wave function up to the normalization constant
\begin{equation}
\Psi=\frac{C_N}{\sqrt{V}} e^{iS} {\cal U}_0  .
 \label{eq:av final_psiD}
\end{equation}
Since the wave function should be a function of $x$ rather than $\tau$, the phases $\phi_1,\phi_2$ are replaced by $k_1 \cdot x$ , and $k_2 \cdot x$ as in the scalar case.

%At the end, it should be mentioned that the solution for ${\cal U}_0$ given above is analogous to the one appearing in the supplemental material of \cite{DiPiazza_2014}. The difference lies in the fact we parameterize the trajectory 	with the proper time $\tau$, while in \cite{DiPiazza_2014} it is parameterized using the light cone coordinate $(t-z)/2$.

\subsection{The matrix expansion convergence}
%In the previous section, approximated solutions to the scalar and spinor wavefunctions were established.
%For the scalar case, one can easily plug the waveunfunction into the original equation and show it obeys it up to a small quantity corresponding to the second order WKB term, seen in Eq.~\eqref{eq:av WKB_cond}. However, for the spinor case, a direct substitution is cumbersome, because of the matrix nature of the Dirac equation (\ref{eq:av Dirac0}). However, the solution for the spinor case has only one more spin terms, fulfilled Eq.~\eqref{eq:av eq_spin}, compared with the scalar case and therefore to demonstrate the convergence, we only need to demonstrate the convergence of this term with respect to Eq.~\eqref{eq:av eq_spin}. Hence, we explicitly demonstrate in this subsection the convergence of the matrix solution for a typical example.

In the previous section, approximate solutions to the scalar and spinor wave functions were established. For the scalar case, the wave function can be directly substituted into the original Klein-Gordon equation to verify that it satisfies the equation up to a small term corresponding to the second-order WKB correction, as shown in Eq.~\eqref{eq:av WKB_cond}. In contrast, for the spinor case, direct substitution into the Dirac equation \eqref{eq:av Dirac0} is more cumbersome due to the matrix structure of the equation. However, the spinor solution differs from the scalar case only by an additional spin-dependent term, which satisfies Eq.\eqref{eq:av Om_eq}. Consequently, demonstrating the convergence of the spinor solution reduces to verifying that this spin-dependent term converges with respect to Eq.\eqref{eq:av Om_eq}. This simplifies the analysis, allowing the convergence properties of the spinor solution to be inferred directly from the behavior of this term and the convergence condition for the scalar case [Eq.~\eqref{eq:av WKB_cond}].

%In the following, we have check our solution of Eq.\eqref{eq:av eq_spin} order by order to show the convergence. In order to do so, the parameters of the laser beams and particle beams have to be specified first. The working point was chosen according to the following requirements. (a) The validity conditions of the classical trajectory solution (\ref{eq:av condP1}-\ref{eq:av condP2}) are satisfied; (b) the electric field amplitude of the waves, $\sim m \xi_1 \omega, m \xi_2 \omega$ is bellow the Schwinger field value, so pair production is inhibited \cite{Schwinger_1951}; (c) The WKB validity criterion (\ref{eq:av WKB_cond}) is fulfilled.

\begin{table} [b]
  \caption{The value of the residual term
  %The norm of residue
  of the spinor equation (\ref{eq:av Om_eq}) as a function of the expansion order n included in the wave function. The calculation parameters are $\xi_1=2,\xi_2=0.5,\omega=0.05m$. The result is normalized to $m$ and averaged over time and space.}
  \centering
    \begin{tabular}{|c|c|c|c|c|c|}
      \hline
       n & 1 & 2 &3& 4  & 5 \\
      \hline
        $\lvert \cal{R} \rvert$ &$4.7\times 10^{-2}$ &$2.4 \times 10^{-3} $&$3.1 \times 10^{-5}$& $7.6\times 10^{-7}$ &$4.8\times 10^{-9}$ \\
      \hline
    \end{tabular}
    \label{Tab:table1}
\end{table}

In the following, we verify the solution of Eq.\eqref{eq:av Om_eq} in the form of Eq.~(\ref{eq:av recursive1}) order by order to demonstrate convergence.
%To achieve this, the parameters of the laser and particle beams must first be specified.
Firstly, we specify the parameters of the laser and particle beams, which should fulfill
%The working point is chosen based on
the following requirements:
\begin{itemize}
\item For a significant quantum effect,
%In order to observe a significant effect on the quantities presented in this section,
high laser frequency and amplitudes are favourable. Hence, a Free Electron Laser facility operating in the x-ray regime (XFEL) is the best candidate. The chosen laser frequency is reachable in present  XFEL facilities and the amplitudes are higher by about an order of magnitude but might be accessible in the next generation ones.
\item The validity conditions for the classical trajectory solution, Eqs.(\ref{eq:av condP1})-(\ref{eq:av condP3}), are satisfied.
\item The WKB validity criterion, Eq.~(\ref{eq:av WKB_cond}), is fulfilled.
\item The electric field amplitudes of the waves, $m \xi_1 \omega$, and $m \xi_2 \omega$, remain below the Schwinger field threshold, ensuring that pair production is suppressed \cite{Schwinger_1951}.
\end{itemize}

Table \ref{Tab:table1} presents the average over time and space of the value of the residual term
%residue's norm
 $\lvert \cal{R} \rvert=\sqrt{\mathcal{R}^{\dagger} \mathcal{R}}$, as defined in Eq.~\eqref{eq:av residue}, for several orders of the approximate expansion. The results are normalized to $m$, and indicate a rapid convergence, requiring only a few terms to achieve high accuracy.

Before using the wave function to calculate some physical quantities, let us briefly compare the wave function obtained above with the ones presented in \cite{Hu_2015,King_2016}. In \cite{Hu_2015} an approximate solution for the wave function of spinless particles in counterpropagating waves are derived, as well as its WKB limit. Since the authors assume that the angle between the particle propagation and the beams axis is $\sim 1$, there is no overlap between their solution and the one presented here, where this angle is assumed to be very small. Nevertheless, the wave function presented here is more favorable for rate calculation for two reasons. First, the validity condition required in \cite{Hu_2015} are much more restrictive, see Eqs. (41), (55)-(58) there. Second, their solution is given in terms of an infinite series, rendering the rate calculation very cumbersome. On the other hand, in \cite{King_2016} a particle propagating on the beams axis is considered, and a classical solution and the corresponding WKB wave function are derived. The momentum $P_z$ takes then the form of an elliptic integral. In case Eqs.~(\ref{eq:av condP1})-(\ref{eq:av condP3}) are fulfilled, the integrand may be Taylor expanded, reducing to the above given expression Eq.~(\ref{eq:av final_LI}) with $p_x=0$. However, the nontrivial relation between the asymptotic and the average momentum, derived at the end of subsection~\ref{subsec:traj} in our paper, is not discussed there. Moreover, their solution cannot be used to rate calculation even if the incoming particle transverse momentum is vanishing.
%satisfies $p_{\bot}=0$.
The reason is that in this case the outgoing particle will acquire transverse momentum, thus going beyond the approximation validity. The wave function given here, however, applies also for nonnegligible transverse momentum
%$p_{\bot} \neq 0$
 and therefore is suitable for a rate calculation.

\section{Spin dynamics}
\label{sec:num}

In this section, we use the previously derived WKB wave function to calculate various physical %quantities
characteristics and compare them with possible classical counterparts, highlighting the role of the quantum effects.

% This comparison will highlight some quantum properties of the electron during its dynamics.

\begin{figure}
  \begin{center}
  \includegraphics[width=0.5\textwidth]{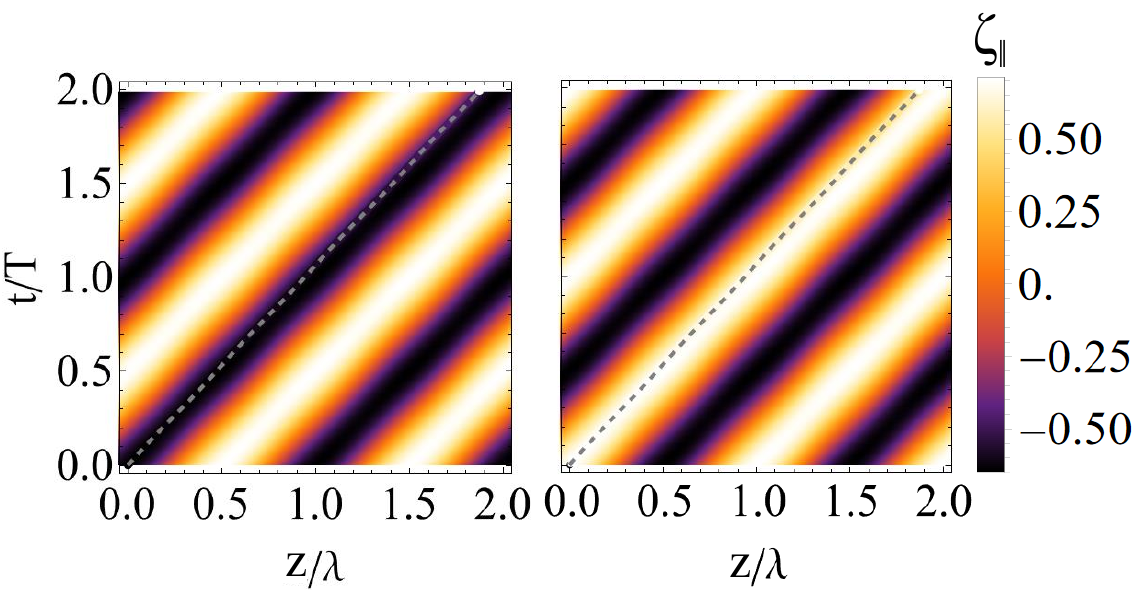}
  \caption{The helicity
  %spin parallel to the particle's motion
  $\zeta_{\parallel}$. The right sub-figure designates the case of initial spin-up and the left one the case of initial spin-down. The gray dashed line is the classical trajectory. The parameters are $\xi_1=\xi_2=1,\,\omega=0.5m,\bar{\varepsilon}=5m,\,p_{\bot}=0$.}
  \label{fig:S_2D}
  \end{center}
\end{figure}

\begin{figure}
  \begin{center}
  \includegraphics[width=0.45\textwidth]{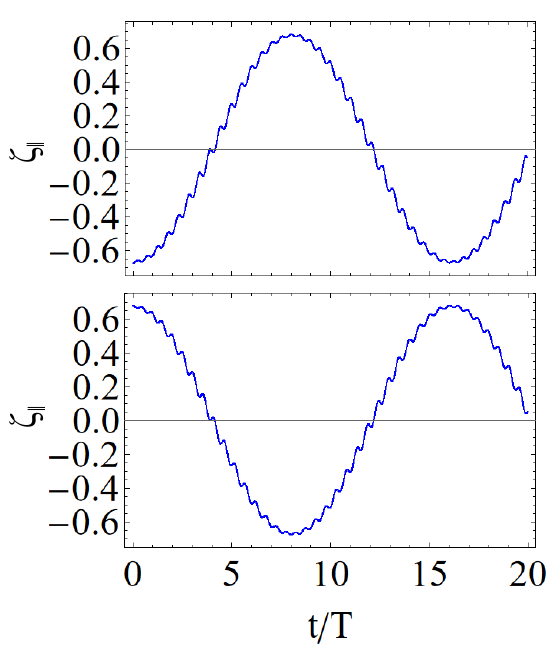}
  \caption{ The helicity
  %spin $\zeta_{\parallel}$
  along the classical trajectory for the spin up (upper Panel) and spin down (lower Panel) cases, respectively. Calculation parameters: $\xi_1=\xi_2=1,\,\omega=0.5m, \,\bar{\varepsilon}=5m,p_{\bot}=0$.}
  \label{fig:S_1D}
  \end{center}
\end{figure}
We start by examining the spin of the electron, as it has no classical counterpart and therefore emphasizes the purely quantum effects of the electron's dynamics.
It can be defined as
\begin{equation}
  \label{eq:av spin_def}
  {\zeta}_{\mu} \equiv \bar{\psi} \gamma_5 \gamma_{\mu}\psi.
\end{equation}
It should be mentioned that the expectation value of the spin calculated from a WKB wave function also obeys the familiar Bergmann-Michel-Telegdi (BMT) equation, as was generally proven in Ref.~\cite{Rafanelli_1964}.
%\begin{equation}
%  \label{eq:av BMT}
%  \frac{d \zeta^{\mu}}{d \tau} = \frac{e}{m} F^{\mu \nu} \zeta_{\nu}.
%\end{equation}
In the instantaneous rest frame of the electron, the 4-pseudovector spin is expressed as $(0,\pmb{\zeta})$. In many high-energy particle reactions,
%calculations \cite{},
the helicity (namely, the component of $\pmb{\zeta}$ parallel to the momentum $\pmb{P}$) is of particular interest
\begin{equation}
  \label{eq:av spin_||}
  \zeta_{\parallel} =\frac{m}{\lvert \pmb{P} \rvert} \zeta_0.
\end{equation}

%In this section we demonstrate the dynamical behaviour of several physical quantities which are expected to exhibit a quantum effect. The examined quantities are $\zeta_{\parallel}$, the parallel component of the spin (with respect to the motion's direction), as well as the components of the stress-energy tensor (namely the energy and momentum densities).
%The calculation parameters $\xi_1=\xi_2=1,\bar{\varepsilon}=5m,\omega=0.5m, p_{\bot} = 0$ according to the criteria detailed in the previous subsection.
In the calculations below, we choose the spin to be initially oriented along the $x$-axis, perpendicular to the initial average velocity of the particle ($z$ axis). This configuration is selected because it maximizes the value of $\zeta \cdot B$, amplifying the spin's influence on the particle's motion.

\begin{figure}
  \begin{center}
  \includegraphics[width=0.45\textwidth]{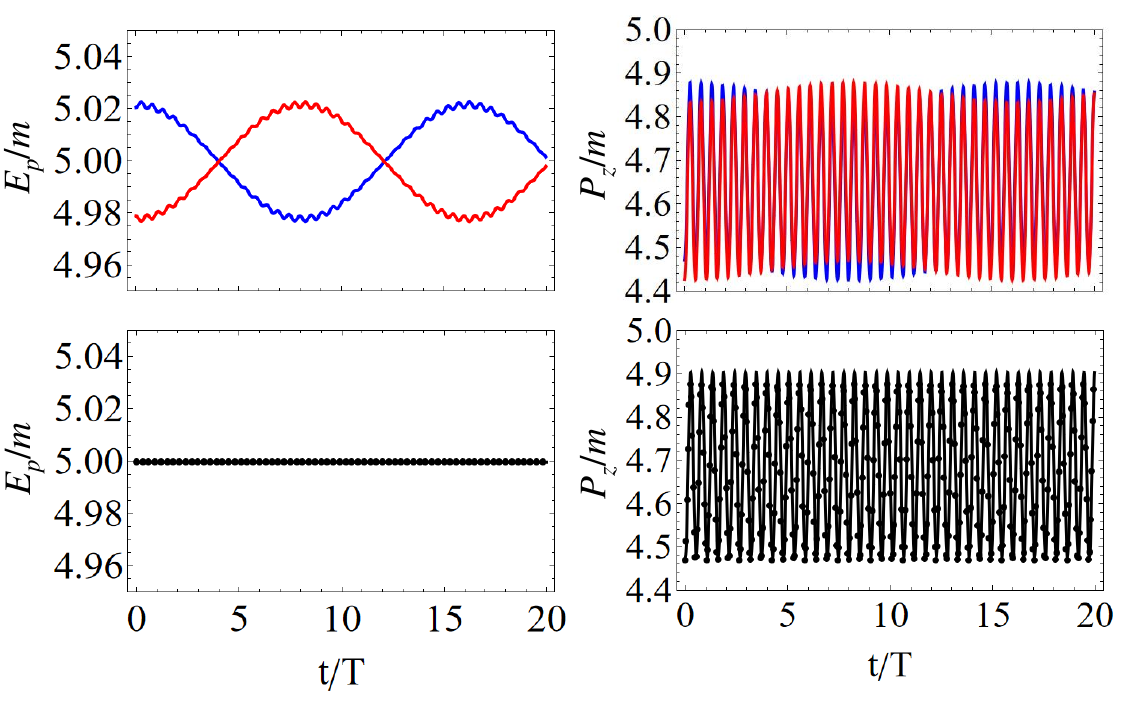}
  \caption{The energy (left column) and momentum (right column) along the classical trajectory. The upper figures stand for the spin up (blue) and spin down (red) cases. The lower figures show the spin-averaged value (black curve) and classical value (black dots) respectively. Calculation parameters: $\xi_1=\xi_2=1,\omega=0.05m, \bar{\varepsilon}=5m,p_{\bot}=0$.}
  \label{fig:Qmomentum}
  \end{center}
\end{figure}
\begin{figure}[b]
  \begin{center}
  \includegraphics[width=0.45\textwidth]{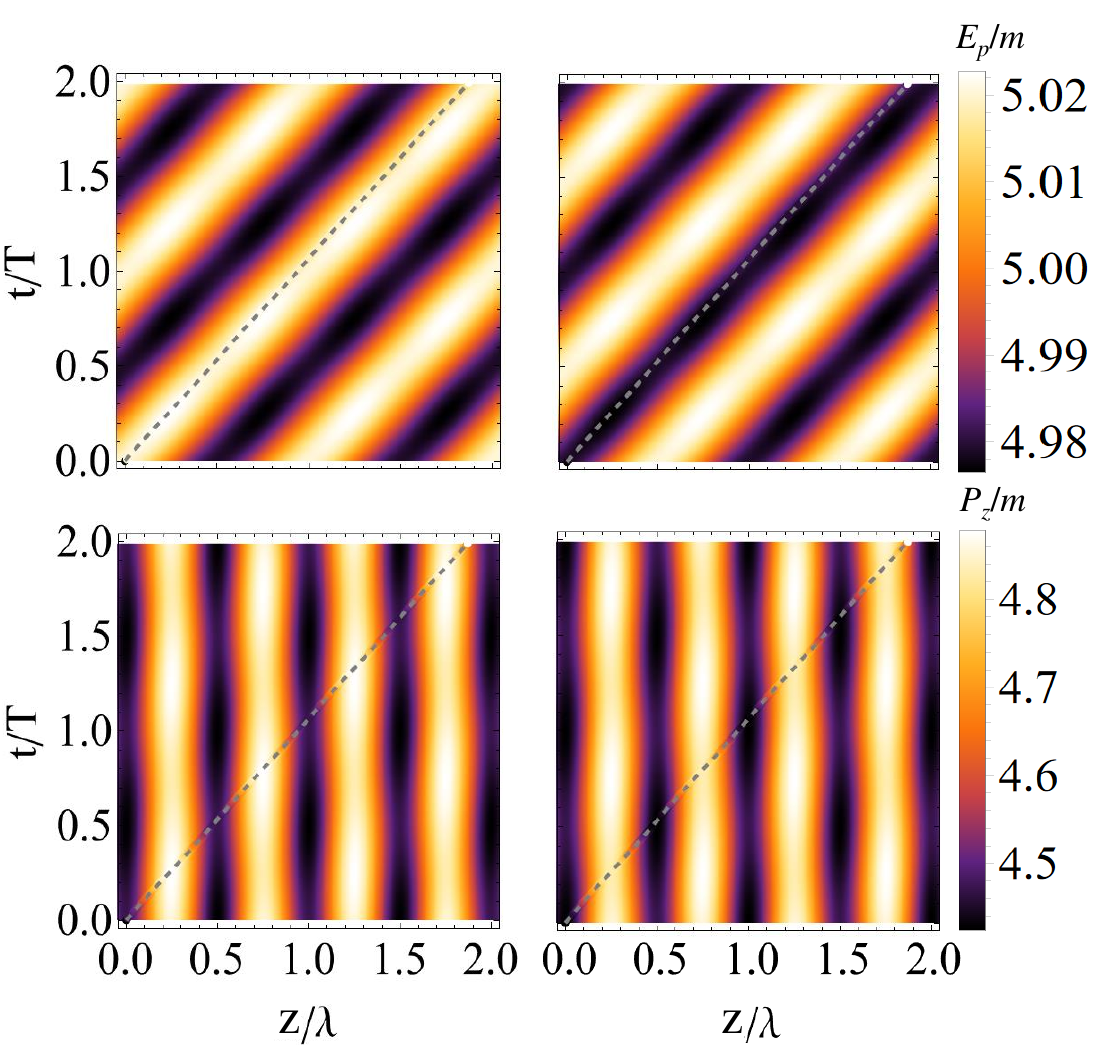}
  \caption{ The energy and momentum corresponding to the wave function. The right sub-figure designates the spin up case and the left one the spin down case. Calculation parameters: $\xi_1=\xi_2=1,\omega=0.05m, \bar{\varepsilon}=5m,p_{\bot}=0$.}
  \label{fig:E_2D}
  \end{center}
\end{figure}
\begin{figure}[b]
  \begin{center}
  \includegraphics[width=0.45\textwidth]{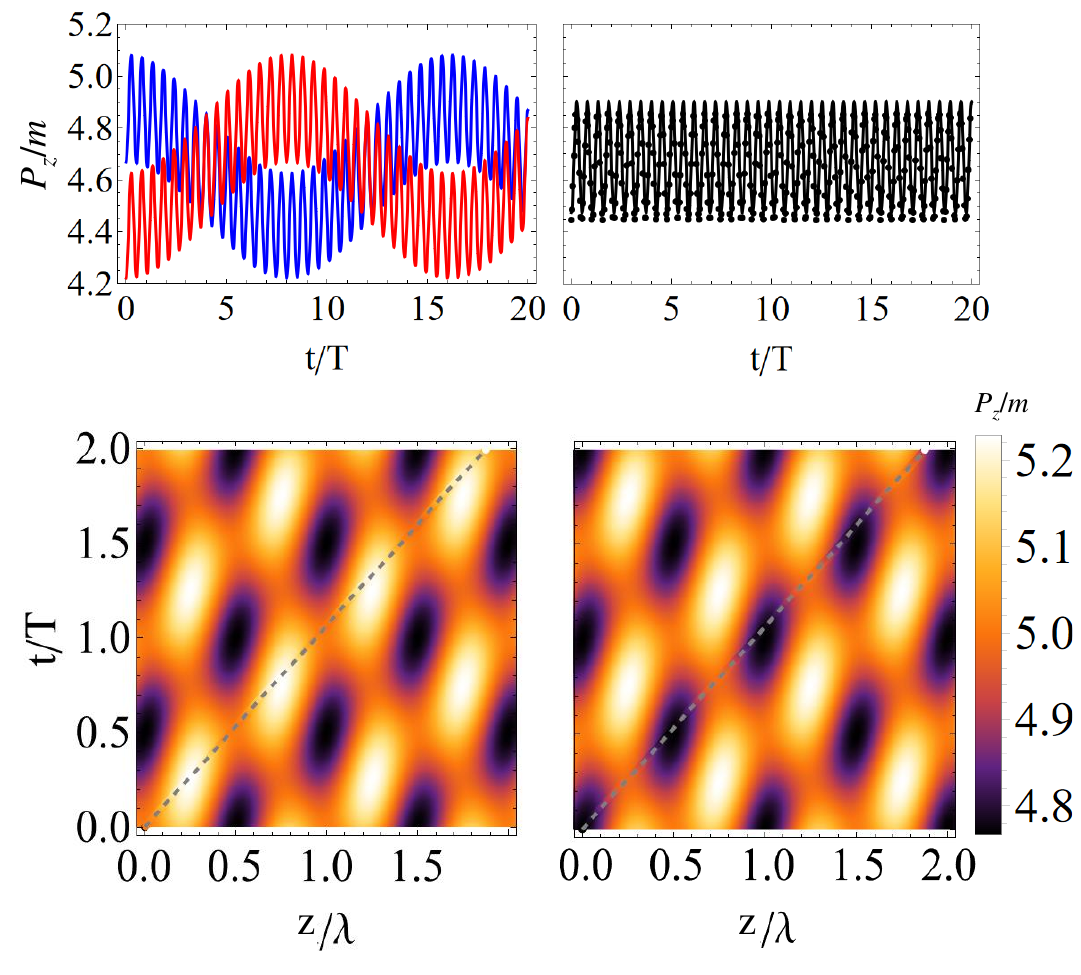}
  \caption{ The momentum corresponding to the wave function for high frequency lasers. The lower right sub-figure designates the spin up case and the lower left one the spin down case. 
The upper left sub-figure represents the spin up (blue) and spin down (red) cases along the classical trajectory.  
 The upper right sub-figure shows the spin-averaged value (black curve) and classical value (black dots). Calculation parameters: $\xi_1=\xi_2=1,\omega=0.5m, \bar{\varepsilon}=5m,p_{\bot}=0$.}
  \label{fig:P2_2D}
  \end{center}
\end{figure}

The wave function obtained in the previous section is a function of the 4-vector $x_{\mu}$. Assuming the axis of average motion is $z$, the various phases appearing in the wave function's expression depend on $z,t$. Therefore, a better understanding of the spin dynamics arises when one examines the 2D plots (see Fig.\ref{fig:S_2D}) of the spin as a function of the temporal and spatial coordinates. From the figure we can see that the spin of the electron oscillates in a wide range, $\pm 0.65$.
In order to explain the pattern we recall that according to our analytical solution, 3 types of oscillations are to be expected, corresponding to the three phases $k_1 \cdot x, k_2 \cdot x, \Delta k \cdot x $. The large amplitude ones in the spin arise from the $k_1 \cdot x$ term. For this reason, they take the form of diagonal lines, corresponding to the copropagating beam.
The velocity of the electron is smaller than the speed of light, which makes its classical trajectory (gray dashed lines in the figure) lag slowly behind. This fact gives rise to the slow oscillations with frequency $\omega_1$ [see definition in Eq.~\eqref{eq:av om12}], seen also in Fig.~\ref{fig:S_1D}. The imprint of the second laser pulse is much smaller and is therefore not visible in these plots.

%\textcolor{blue}{The classical trajectory is presented on top of these 2D maps. The straight lines correspond to the co-propagating laser. One may see that the particles  (as its velocity is smaller than the speed of light) giving rise to the slow oscillation with frequency $\omega_1$. The imprint of the second pulse is much smaller and is therefore not distinguishable in these plots.  }

As is  well known, in the realm of the WKB approximation, the particle's free propagation (without taking into account QED processes) approximately follows the classical trajectory $x_{\mu} \left( \tau \right)$. Therefore, we now examine in Fig.\ref{fig:S_1D} the spin $\zeta_{\parallel}$ along the classical trajectory as a function of the laboratory time. Two typical frequencies are observed in the spin dynamics, corresponding those appearing in the classical motion (\ref{eq:av om12}). The larger frequency $\omega_2$ corresponds to the counter-propagating beam through $k_2 \cdot \bar{P}$ and the small frequency $\omega_1$ relates to the co-propagating beam through $k_1 \cdot \bar{P}$. The ratio between these two frequencies scale as $\gamma_*^2$ and is enhanced by orders of magnitude for  ultrarelativistic particles.
%may be orders of magnitude high for ultra-relativistic particle.

Secondly, let us write down the symmetric stress-energy for the electron
\begin{equation}
  \label{eq:av stress_energy}
  T_{\mu \nu} \equiv \frac{1}{4}\bar{\psi} \left[ \gamma_{\mu}\overrightarrow{D_{\nu}} + \gamma_{\nu}\overrightarrow{D_{\mu}}
                                                - \gamma_{\mu}\overleftarrow{D_{\nu}} - \gamma_{\nu}\overleftarrow{D_{\mu}} \right] \psi,
\end{equation}
with $D_{\mu} \equiv i \partial_{\mu}-eA_{\mu}$. The left (right) arrow designates an operator acting to the left (right). The $0^{th}$ component of this tensor, $T_{0 \mu }$, is equivalent to the momentum density. Therefore, dividing by the probability density yields the quantum counterpart of the classical momentum, $P_{\mu}=T_{0 \mu }/\psi^{\dagger} \psi$.

The energy and momentum along the classical trajectory are shown in Fig.\ref{fig:Qmomentum}. The black line stands for the spin averaged quantities and the black dots to the classical value. One can observe, as a sanity check, that these two curves coincide, as expected. The spin-dependent quantities exhibit more complex behaviour as compared to the spin-averaged ones. For example, the spin-averaged energy is constant whereas the spin-dependent energy manifests signatures of the frequencies $\omega_1$, and $\omega_2$ discussed above.

Again, the corresponding 2D plots are presented in Fig~\ref{fig:E_2D} to give a broader picture of the dynamics. The energy plot is analogous to the spin component $\zeta_{\parallel}$ discussed earlier. As for the momentum, the oscillations may be separated to a semi-classical part, originating from the $\Delta k \cdot x$ term, and a spin-dependent quantum part, associated with the $k_1 \cdot x, k_2 \cdot x$ terms. In Fig~\ref{fig:E_2D}, the dominant contributions is the semi-classical one, and hence one observes vertical lines (since $\Delta k \cdot x = 2 \omega z$ is time-independent).
In case $\omega$ is increased, the quantum part becomes more manifest as can to be seen in  Fig~\ref{fig:P2_2D}. Namely, the spin-dependent oscillations are enhanced (right column) and the pattern in the 2D plot is modified accordingly (left column). Such high values of the frequency ($\omega=0.5m$) are not currently feasible but are useful in order to demonstrate the interplay between the two contributions .

\section{Discussion and conclusions}
\label{sec:con}
In this paper, the quantum dynamics of a scalar/fermion particle in the  CPW setup is explored. Assuming the particle is moving with relativistic velocity and a small angle with respect to the CPW axis, the frequency of the co- and counter-propagating beams in the particle's frame significantly differ from each other. As a result, the system is far from resonance and the Lorentz equation can be approximately integrated analytically.
With the aid of the classical trajectory, the Hamilton-Jacoby equation for the classical action is integrated as well. This equation also corresponds to the zeroth-order term of the WKB expansion. Afterwards, the equation describing the first-order WKB term was addressed utilizing the characteristics method. For the scalar case, the integration is exact. For the spinor case, this equation does not admit an exact solution, since the ODE coefficients are matrices rather than c-numbers and therefore do not generally commute with themselves for different times.
A unique expansion was used, whose terms are related to each other recursively through a set of ODE's.
Fortunately, given the assumption required in the realm of our classical solution approximation, this set of equations is solved, allowing one to write down general expression for the expansion terms and show their amplitude is decreasing. The validity condition was discussed by evaluating the order of magnitude of the neglected term in the equation. The bottom line coincides with the estimation of Baier and Katkov. Accordingly, the semiclassical approach is adequate as long as the particle energy greatly exceeds the typical frequency of the particle's motion induced by the external field.

The final result of the derivation, namely the particle's wave function, paves the way to various research possibilities related to the quantum dynamics of a particle in the presence of a CPW background. For instance, it may be applied to calculate the Schwinger pair production rate for a standing wave. It is well known that this QED process takes place only for electromagnetic configuration with $\mathcal{F} \equiv \frac{e}{m^2}\sqrt{E^2-B^2} \sim 1$. The simplest way to realize it in the laboratory is the CPW configuration. At present, however, analytical research of this topic is restricted to the pure electric field configuration \cite{Hebenstreit_2009}. It should be mentioned that since our wave function is valid for certain momentum values, only part of the phase space may be reachable. Nevertheless, it can still be valuable for physical insights as well as for benchmarks with extensive numerical models such as the Wigner formalism \cite{bialynicki1991phase,hebenstreit2011particle}.

Furthermore, this wave function is suitable for calculations of strong-Field QED scatterings, such as nonlinear Compton and Breit-Wheeler (first-order processes) as well as higher order ones (e.g. trident process, mass correction etc). The higher order processes were out of reach until now in the absence of an appropriate wave function. the first order processes may be also obtained within the Baier-Katkov formalism (see for instance \cite{Lv_2020b} regarding the non-linear Compton in CPW), and a detailed comparison would be of fundamental interest. Another intriguing perspective the above presented solution is the spin dynamics of the particle in CPW configuration.

\section*{Acknowledgement}
QZL and ER contributed equally to the work, to numerical and analytical calculations, respectively. The authors are grateful to K.Z. Hatsagortsyan and C.H. Keitel for their support and valuable comments. ER wishes to thank A. Zigler for fruitful discussions.

\bibliography{cpw_wkb}

\end{document}